\documentclass[11pt,twoside]{article}

\usepackage{asp2004}
\usepackage{epsf}
\usepackage{psfig}
\usepackage{lscape}

\begin{document}
\title{Weighing Black Holes in High-Redshift SCUBA Galaxies}   
\author{D.M. Alexander}   
\affil{Department of Physics, Durham University, Durham DH1 3LE, UK}    

\begin{abstract} 

  Deep SCUBA surveys have uncovered a population of dust-enshrouded
  star-forming galaxies at $z\approx$~2. Using the ultra-deep 2~Ms
  {\it Chandra} Deep Field-North survey we recently showed that a
  large fraction of these systems are also undergoing intense
  black-hole growth. Here we provide further constraints on the
  properties of the black holes in SCUBA galaxies using the virial
  black-hole mass estimator. We show that typical SCUBA galaxies are
  likely to host black holes with $M_{\rm
    BH}\approx10^{7}$--$10^{8}$~$M_{\odot}$ which are accreting at, or
  close to, the Eddington limit. These results provide qualitative
  support for our earlier conclusion that the growth of the black hole
  lags that of the host galaxy in these massive ultraluminous
  galaxies.

\end{abstract}



\section{Tracing the Growth of Massive Galaxies}

Deep surveys over the past decade have revolutionised our
understanding of the high-$z$ Universe and mapped the history of star
formation out to $z\approx$~6 (e.g.,\ \citealt{madau98}). Over the
same timescale, the finding that every massive galaxy hosts a central
super-massive black hole (SMBH) with a mass proportional to that of
its spheroid (e.g.,\ \citealt{trem02}) has raised the surprising
possibility that the growth of galaxies and their SMBHs are
concurrent, despite nine orders of magnitude difference in size scale.

The most active sites of star formation at high redshift are
identified with (sub-)millimetre emitting galaxies (SMGs), such as
those detected in deep SCUBA surveys (e.g.,\ \citealt{smail02}).  The
estimated star-formation rates of these systems are large enough to
form a massive galaxy of $\approx10^{11}$~$M_{\odot}$ in just
$10^{8}$~yrs (e.g.,\ \citealt{chap05}; \citealt{greve05}).  After
intense multi-wavelength follow-up observations it is clear that these
SMGs are already massive at their average redshift of $z\approx$~2 and
are sufficiently gas rich to maintain a vigorous level of star
formation for a further $\approx5\times10^{7}$~yrs (e.g.,\
\citealt{swin04}; \citealt{tecza04}; \citealt{tacc06}).  These results
suggest that SMGs pinpoint the formation of local $>M_{*}$ early-type
galaxies. Given the apparent connection between the growth of galaxies
and their black holes, what can be understood about the growth of
black holes in SMGs?

\section{The Growth of Black Holes in SMGs}

Using ultra-deep {\it Chandra} observations (the 2~Ms {\it Chandra}
Deep Field-North survey) we recently showed that $\approx$~40\%
(perhaps all) of the bright ($f_{\rm 850\mu m}>4$~mJy) SMG population
host heavily obscured Active Galactic Nuclei (AGN) activity (e.g.,\
\citealt{alex05a}; \citealt{alex05b}). The energetics of the AGN
activity was found to be (typically) too low to explain the huge
bolometric output of these objects, which is almost certainly
dominated by star formation (see also, e.g.,\ \citealt{chap04};
\citealt{ivison04}).  However, the large AGN fraction implies that the
SMBHs are growing almost continously throughout these periods of
vigorous star formation (i.e.,\ $>$~40\% duty cycle of SMBH fueling).
This almost continuous SMBH growth suggests that there is an abundance
of available fuel, hinting that the accretion may be occuring at, or
close to, the Eddington limit.  Although hypothetical, this picture is
in good agreement with direct predictions for the growth of SMBHs in
SMG-like systems (e.g.,\ \citealt{dima05}; \citealt{king05};
\citealt{gran06}). Under the assumption of Eddington-limited accretion
we showed that the masses of the SMBHs in typical SMGs are
$<10^{8}$~$M_{\odot}$ (\citealt{alex05a}).  Fig.~1 shows the mass
accretion rate for SMGs hosting AGN activity and a comparison sample
of $M_{\rm B}<-24$ quasars.

Utilising deep rest-frame optical--near-IR observations we also
directly estimated the stellar masses of these X-ray detected SMGs,
finding that their Eddington-limited SMBH masses are $\approx$~1--2
orders of magnitude below that measured for comparably massive
galaxies in the local Universe (\citealt{borys05}).  This result
suggests that either (1) the growth of the SMBH lags that of the host
galaxy in SMGs, or (2) the SMBHs in SMGs are accreting at
sub-Eddington rates.  The former is in stark contrast with the growth
of SMBHs estimated for high-$z$ quasars and radio galaxies (e.g.,\
\citealt{walter04}; \citealt{mclure06}) while the latter is in
conflict with that predicted by models for the growth of SMG-like
systems with an abundance of available fuel. The aim of this brief
contribution is to further constrain the properties of the black holes
in SMGs.

\begin{figure}[!t]
\plotone{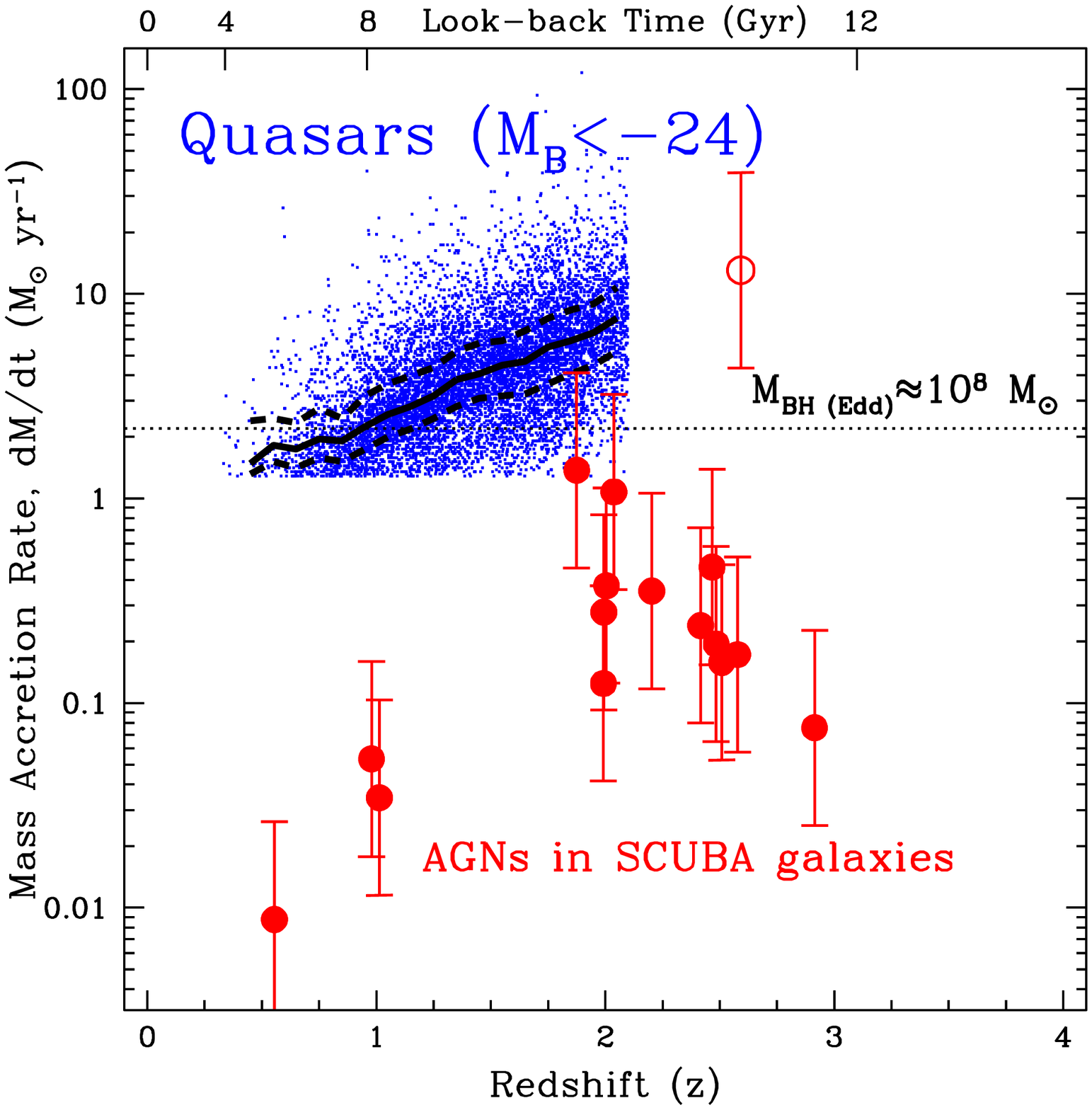}
\caption{Black-hole mass accretion rate vs redshift for the SMGs
  (circles) and a comparison sample of $M_{\rm B}<-24$ quasars (blue
  dots). The SMGs are split into typical SMGs with ultra-deep X-ray
  constraints from \cite{alex05a} (solid circles) and an X-ray
  detected SMG with broad H$\alpha$ from \cite{swin04} (open circle);
  the mass accretion rates for the SMGs are determined following the
  luminosity dependent bolometric correction factors of \cite{marc04}
  and are consistent with \cite{alex05a}. The quasar data is from
  \cite{mclure04}. The dotted line indicates the approximate
  Eddington-limited mass accretion rate for a $10^{8}$~$M_{\odot}$
  black hole ($M_{\rm BH (Edd)})$ while the solid and dashed curves
  show the median and interquartile ranges for the quasars,
  respectively.  Typical SMGs have mass accretion rates approximately
  an order of magnitude below those of coeval quasars, indicating
  smaller black-hole masses for comparable Eddington accretion rates
  ($M_{\rm BH}<10^{8}$~$M_{\odot}$). Adapted from \cite{alex05a}.}
\end{figure}

\begin{figure}[!t]
\plotone{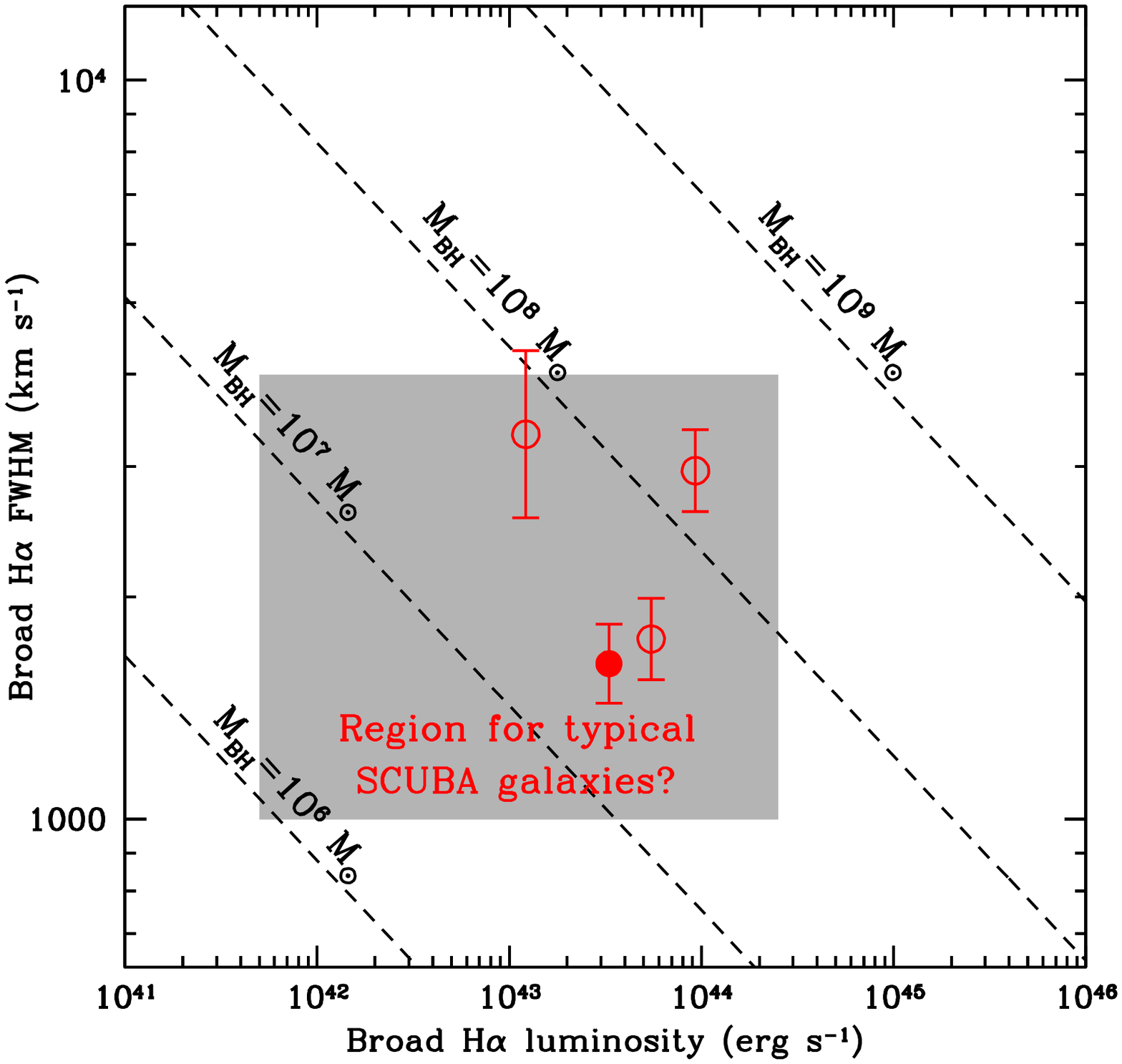}
\caption{H$\alpha$ emission properties of the four SMGs with broad
  H$\alpha$ from \cite{swin04}. The filled circle indicates the SMG
  with ultra-deep X-ray constraints from \cite{alex05a} and the open
  circles indicate the other SMGs from \cite{swin04}. The dashed lines
  show the relationship between $M_{\rm BH}$ and the broad H$\alpha$
  emission-line properties on the basis of the virial black-hole mass
  estimator of \cite{greene05}. The shaded region shows the likely
  range of properties for typical SMGs without detected broad
  H$\alpha$ emission ($M_{\rm
  BH}\approx10^{7}$--$10^{8}$~$M_{\odot}$); see \S3.}
\end{figure}

\begin{figure}[!t]
\plotone{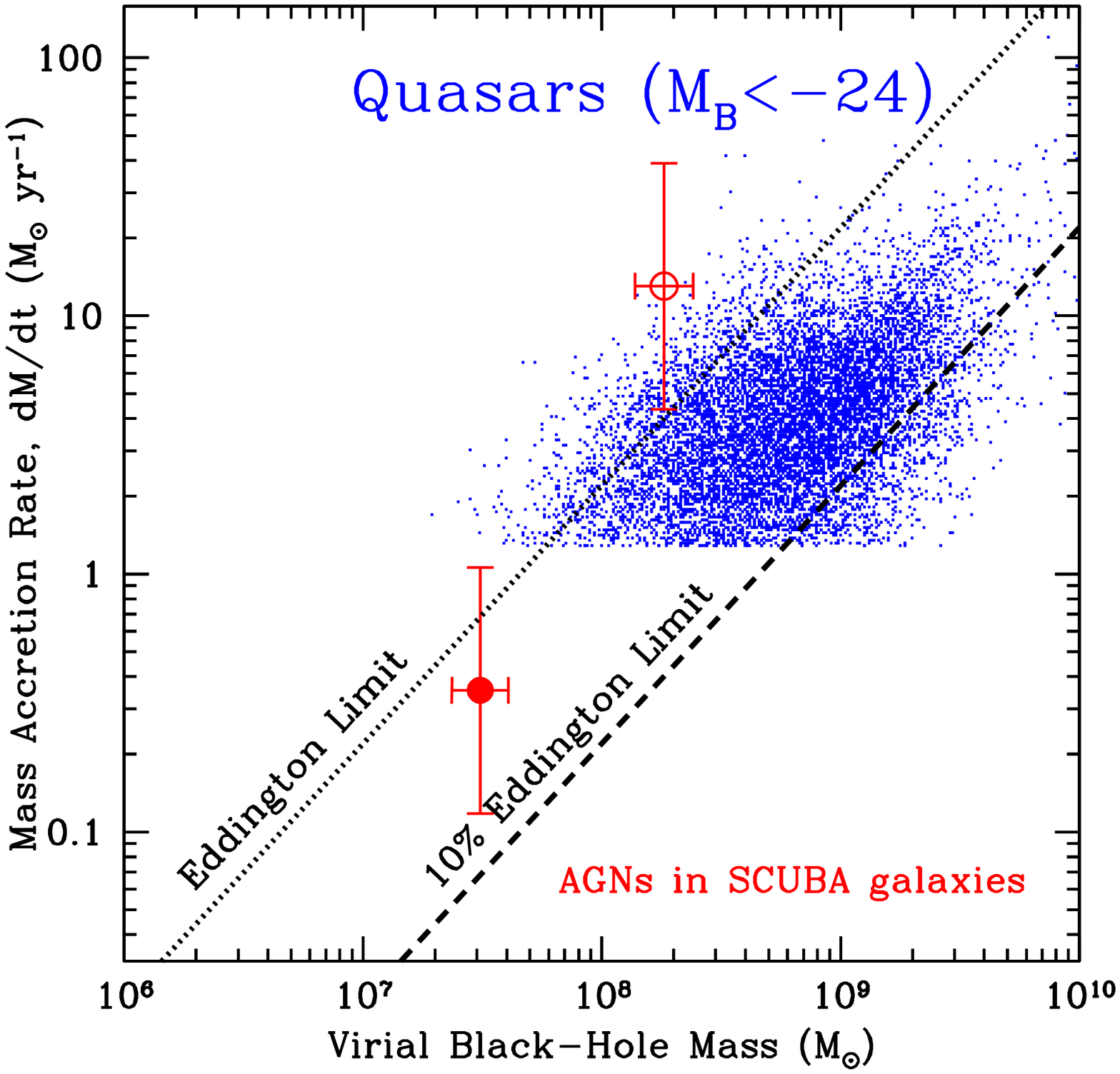}
\caption{Mass accretion rate vs black-hole mass of the two X-ray
  detected SMGs with broad H$\alpha$ (circles) and a comparison sample
  of $M_{\rm B}<-24$ quasars (blue dots); see Figs.~1 \& 2 for further
  definitions. The dotted and dashed lines indicate the relationship
  for Eddington-limited accretion and accretion at 10\% of the
  Eddington limit, respectively. The SMGs have properties consistent
  with efficient black-hole accretion (i.e.,\ at, or close to, the
  Eddington limit).}
\end{figure}

\section{The Properties of the Black Holes in SMGs}

Black-hole masses ($M_{\rm BH}$) for a small number of galaxies in the
local Universe have been directly measured on the basis of the
velocity dispersion of stars/gas in the vicinity of the SMBH (e.g.,\
\citealt{gebh00}). Black holes cannot be ``weighed'' in the same way
at high redshift due to poorer spatial resolution and lower
signal-to-noise ratio data.  However, the well-established virial
black-hole mass estimator, which works on the assumption that the
broad-line regions (BLRs) in AGN are virialized, provides an
apparently reliable, if indirect, measurement of SMBH masses in
high-$z$ AGN (e.g.,\ \citealt{kaspi00}). Using this technique, the
masses of SMBHs in quasars up to $z\approx$~6.4 have been estimated
(e.g.,\ \citealt{willot03}; \citealt{mclure04}).

The virial black-hole mass estimator is somewhat restrictive for SMGs
since the majority of the sources are heavily obscured (e.g.,\
\citealt{alex05b}). However, the identification of broad H$\alpha$
emission from a small number of SMGs (\citealt{swin04}) provides the
opportunity to estimate SMBH masses for a handful of sources so far.
Here we use the virial estimator of \cite{greene05}, which calculates
$M_{\rm BH}$ solely on the properties of the H$\alpha$ emission line
and reduces potential uncertainties on the luminosity of the AGN
(e.g.,\ contaminating emission from the host galaxy or an
accretion-related jet) when compared to other estimators.

Fig.~2 shows the broad H$\alpha$ properties of the four SMGs from
\cite{swin04} with FWHM(H$\alpha$)$>1000$~km~s$^{-1}$.  These SMGs
have narrow H$\alpha$ emission lines when compared to the quasar
population ($\approx$~2000~km~s$^{-1}$ vs typically
$\approx$~5000~km~s$^{-1}$) which leads to relatively modest SMBH
masses (average $M_{\rm BH}\approx5\times10^{7}$~$M_{\odot}$). Can
these SMBH masses also be considered typical of the overall SMG
population? Although we cannot directly measure the broad H$\alpha$
emission-line properties of the other SMGs in our sample, we can {\it
predict} their broad H$\alpha$ luminosity using the X-ray luminosities
from \cite{alex05b} and the X-ray--H$\alpha$ luminosity relationship
found by \cite{iman04} for low-$z$ galaxies with luminosities
comparable to those of the SMGs (see also \citealt{ward88}). The
predicted broad H$\alpha$ luminosity for typical SMGs is up-to
$\approx3\times10^{44}$~erg~s$^{-1}$, consistent with the maximum
luminosity of the X-ray detected SMGs with broad H$\alpha$ emission.
Therefore, under the reasonable assumption that the broad H$\alpha$
emission line has the same intrinsic width in all SMGs but is
undetected in most systems due to the presence of dust/gas along the
line of site to the BLR (i.e.,\ the unified model of AGN;
\citealt{anto93}), the predicted SMBH mass range for typical SMGs is
$M_{\rm BH}\approx10^{7}$--$10^{8}$~$M_{\odot}$; see Fig.~2.  Indeed,
the masses of the SMBHs in typical SMGs seem unlikely to be much
larger than $\approx10^{8}$~$M_{\odot}$ unless the majority of SMGs
represent a population distinct from those with detected broad
H$\alpha$ and have intrinsically broader H$\alpha$ emission, an
assumption that would fly in the face of current wisdom (e.g.,\
\citealt{page04}; \citealt{stev05}).

What does this result imply for the relative growth of the SMBH and
host galaxy in SMGs? Clearly robust masses are required to provide a
quantitative answer to this question.  However, given the
SMBH--spheroid mass ratio in the local Universe
($\approx1.3\times10^{-3}$; e.g.,\ \citealt{merrit01}), so long as
SMGs have stellar masses $>10^{10}$--$10^{11}$~$M_{\odot}$ then the
growth of their SMBHs are likely to lag that of the host galaxy.
Current photometric and dynamical mass estimates suggest that the host
galaxies of SMGs are of order $\approx10^{11}$~$M_{\odot}$ (e.g.,\
\citealt{swin04}; \citealt{borys05}; \citealt{tacc06}).

Fig.~3 shows the mass accretion rate vs the virial black-hole mass for
the two X-ray detected SMGs with broad H$\alpha$. This plot provides a
measure of how efficiently the SMBHs are growing in SMGs. Even taking
into account the large uncertainties, the SMGs appear to be accreting
at $>10$\% the Eddington limit and the average mass accretion rate is
comparable to the Eddington limit; the finding that AGNs with
similarly narrow BLRs (such as narrow-line Seyfert 1s; e.g.,\
\citealt{osterb85}) are accreting at comparable rates provides
tangential support for this result (e.g.,\ \citealt{mclure04}).
Evidently quantitative conclusions cannot be drawn from such a small
and heterogenus sample, however, these results provide qualitative
support for the models that predict that SMGs are undergoing a rapid
Eddington-limited SMBH growth phase.

\section{Conclusions and Discussion}

We have shown that the black holes in high-$z$ SMGs appear to be
undergoing almost continous growth throughout periods of intense star
formation.  Using the black-hole virial mass estimator we predict that
typical SMGs are likely to host $M_{\rm
  BH}\approx10^{7}$--$10^{8}$~$M_{\odot}$ black holes which are
accreting at, or close to, the Eddington limit. These results provide
qualitative support for (1) the models that predict that SMGs are
undergoing a rapid Eddington-limited SMBH growth phase, and (2) our
earlier conclusion that the growth of the SMBH lags that of the host
galaxy in these massive ultraluminous galaxies.

More detailed analyses of a larger sample of SMGs with high
signal-to-noise near-IR spectroscopy and/or spectropolarimetry in
addition to robust host-galaxy mass estimates will provide a more
quantitative test of these conclusions. Over the next few years, the
expected 10--100\,pc resolution of ALMA will also allow the detailed
dynamics of the gas within the environment of the feeding SMBH to be
probed, providing the potential to directly track the influence of the
SMBH on the dynamics of the gas.

\acknowledgements 

I thank the Royal Society for support, R.~McLure for providing the
quasar data, and my collaborators for allowing me to present this
research: F.~Bauer, A.~Blain, C.~Borys, N.~Brandt, S.~Chapman,
R.~Ivison, I.~Smail, and M.~Swinbank.


\newpage

\end{document}